\documentstyle[12pt]{article} 

\oddsidemargin =  \evensidemargin
\newcommand{\eg}{{\it e.g.~}}
\newcommand{\ie}{{\it i.e.~}} 

\newcommand{\s}{\sigma}
\newcommand{\ap}{\alpha^{\prime}} 
\newcommand{\be}{\begin{equation}}
\newcommand{\ee}{\end{equation}} 
\newcommand{\ba}{\begin{eqnarray}}
\newcommand{\ea}{\end{eqnarray}}

\begin{document} 
 
\def\pct#1{(see Fig. #1.)} 



\begin{titlepage} \hbox{\hskip 12cm ROM2F-97/53  \hfil} 

\begin{center}  
\vspace{0.8cm}
{\Large \bf  A \ Note \ on \ Toroidal \ Compactifications \ of   
\vskip 0.5cm
the \ Type  \ I \ Superstring \ and \ Other \ Superstring   
\vskip 0.6cm
Vacuum \ Configurations \ with \ 16 \ Supercharges}
\vskip 0.8cm

{\large  Massimo BIANCHI } 

\vspace{0.6cm}

{\sl Dipartimento di Fisica, \ \ Universit{\`a} di Roma \ ``Tor Vergata'' \\
I.N.F.N.\ - \ Sezione di Roma \ ``Tor Vergata'', \ \ Via della Ricerca 
Scientifica, 1 \\ 00133 \ Roma, \ \ ITALY} \vspace{0.6cm}
 
ABSTRACT

\end{center}
\vspace{0.8cm}

{We show that various disconnected components of the moduli space of superstring
vacua with 16 supercharges admit a rationale in terms of BPS
{\it un-orientifolds}, \ie type I toroidal compactifications with constant
non-vanishing but quantized vacuum expectation values of the NS$\times$NS
antisymmetric tensor. These include various heterotic vacua with reduced rank,
known as CHL strings, and their dual type II (2,2) superstrings in $D\leq 6$.
Type I vacua without open strings allow for an interpretation of several
disconnected components with $N_V=10-D$. An adiabatic argument relates these
unconventional type I superstrings to type II (4,0) superstrings without
D-branes.  The latter are connected by U-duality in $D\leq 6$ to type II (2,2)
superstrings. We also comment on the relation between some of these vacua and
compactifications of the putative M-theory on unorientable manifolds as well as
F-theory vacua.}

\vspace{0.6cm} 

\end{titlepage} 


\section{Introduction}

The recent astonishing insights on the non-perturbative formulation of
superstrings tend to favor a picture according to which many if not all
consistent vacua emerge as different points on the moduli space of an
underlying theory, commonly termed M-theory. Although the uniqueness of
maximally extended supergravity strongly suggests that vacua with 
32 supercharges in
$D < 10$ be all continuously connected, the
same property does not seem to be shared by vacuum configurations with 16
supercharges. In this note we show that the toroidal compactifications of the
type I superstrings
\cite{bpstor} give a remarkably simple geometric rationale for this
phenomenon. We indeed showed that the moduli space
\footnote{Throughout the
paper, we use the term {\it moduli space} in a broad sense as the space locally
parameterized by the massless scalar fields, often termed {\it moduli}, 
with only
derivative couplings in the low-energy effective lagrangian. These are in
one-to-one correspondence with truly marginal deformations of the conformal
field theory underlying the compactification \cite{dix,bpstor}. 
Note that since there could be more than one disconnected component of the 
moduli space of a theory with 16 supercharges for any given number of 
vector multiplets, one must also consider automorphism symmetries.
Such examples as well as global identifications, known as T- S- and U-dualities
\cite{send}, will be addressed in Sections 3 and 4.} of type I vacua with
16 supercharges, the maximum allowed for this kind of theories in perturbation
theory, includes several disconnected components that can be discriminated
according to the rank of the gauge group or {\it equivalently}  the rank of the
constant non-vanishing expectation value of the NS$\times$NS antisymmetric
tensor \cite{bpstor}. More explicitly, the Chan-Paton (CP) multiplicity,
\ie the number of D9-branes needed to soak up the R$\times$R charge of the
O-planes introduced by the worldsheet parity $\Omega$-projection \cite{as,
bssys, cjp}, turns out to be  
\be N_{CP} = 32 \times 2^{-b/2}
\ee where $b$ is the rank of the constant non-vanishing expectation value of
the internal NS$\times$NS antisymmetric tensor $B_{ij}$ and
is always an even integer. In the unoriented closed-string spectrum of type I
compactifications on $T^d$, the
$d$ vectors from  the mixed components $G_{\mu i}$ of the metric in the
NS$\times$NS sector and the $d$ vectors from  the mixed components $\widetilde
B_{\mu i}$ of the antisymmetric tensor in the R$\times$R sector combine to give
$d$ graviphotons and $d$ vectors in as many vector multiplets. In the presence
of generic Wilson lines in the Cartan subalgebra of the
CP group, the total number of vector multiplets is $N_V=d+ N_{CP}/2$ and the
gauge group for un-conventional toroidal orientifolds with non-vanishing
NS$\times$NS antisymmetric tensor, or BPS\footnote{Here and in
most of the following, BPS does not stand for Bogomolny, Prasad and 
Sommerfield.}
{\it un-orientifolds} for brevity, turns out to be $U(1)^{d+N_V}$
\cite{bpstor}. Further rank reduction may be induced by Wilson lines that lie in
$O(32)$ but not in 
$SO(32)$ \cite{bpstor} or by a modified $\Omega$-projection either in the
closed-string sector \cite{dp,gepner} or in the open-string sector \cite{bstw,
bpstor}. The latter was associated to {\it open-string discrete Wilson lines}
in \cite{bstw} but it is more appropriate to associate it to {\it open-string
discrete torsion}, since it results from an ambiguity in defining the 
M\"obius-strip $\Omega$-projections of the {\it twisted sector} \cite{as} of some
un-orientifolds, in full analogy with the ambiguity that leads to discrete
torsion in some target-space orbifolds \cite{vwdt}.  

Notice that, contrary to naive expectation, a non-vanishing expectation
value of the NS$\times$NS antisymmetric tensor may be perfectly compatible 
with the
$\Omega$-projection.  Indeed, although the {\it fluctuations} of the
NS$\times$NS antisymmetric tensor are projected out of the perturbative
unoriented closed-string spectrum, a properly quantized vacuum expectation
value, \ie one for which $2B$ belongs to the integer cohomology of the
compactification manifold, is consistent with the projection $B\rightarrow -B$
up to a generalized Peccei-Quinn (PQ) shift \cite{bpstor,sese}.

Moreover, the systematic construction \cite{as,bssys} of rational open-string
models strongly motivated the discovery of the BPS
un-orientifolds \cite{bpstor} since it often implies the presence of a
non-vanishing but quantized NS$\times$NS antisymmetric tensor
\cite{gepner,bstw,bpsmin,fpss}. In retrospect, one may thus ascribe the very
existence of disconnected components in the moduli space of superstring vacua
with 16 supercharges to the consistency of the systematic construction
\cite{bssys} of type I superstrings from left-right symmetric type II theories
\cite{as}.

In the heterotic case, similar models with reduced rank, known as CHL strings
\cite{chl}, were found long after the appearance of \cite{bpstor}. The
CHL models, originally constructed in terms of free world-sheet fermions,
display world-sheet current algebras that are typically realized at higher
level. Several CHL models admit an interpretation as compactifications on
asymmetric orbifolds \cite{sv,kk,nsv}, that allows the identification of their
moduli space \cite{cp}. The complete list of possible $N_V$ 
can be found in \cite{cp}. In view of the conjectured duality between
heterotic and type I superstrings in $D=10$ \cite{witdyn,dht}, that has been
tested further in toroidal compactifications with maximal rank \cite{pw}, it is
natural to conjecture that BPS unorientifolds be dual to CHL models
with the same number of vector multiplets. For instance, in
$D=8$ BPS unorientifolds  are expected to be dual to CHL models with
$N_V=18,10$; in $D=6$ to CHL models with
$N_V=20, 12, 8$; in $D=4$ to CHL models 
with $N_V=22, 14, 10, 8$. Recently some of the CHL models have
been given a gauge theoretic interpretation as toroidal compactifications with
non-commuting Wilson lines
\cite{lmst}. In order to sharpen the above duality conjectures, in
Section 3 we offer the type I counterpart of the mechanism of rank reduction. It
is remarkable that an open-string derived concept, such as the non-commutativity
of the Wilson lines, can be given a closed-string interpretation in terms of a
quantized NS$\times$NS tensor. This mechanism may have far reaching consequences
in other important applications of open-string theories \cite{ap}.
 
Further support to the above conjectured duality can be given by matching
the massless spectra of the two theories at points of enhanced gauge symmetry.
To this end, in Section 3 we derive  the
genus-one partition function of the CHL model with $E(8)$ current algebra at
level $k=2$. A proper choice of Wilson lines breaks $E(8)$ to $SO(16)$ or to
other non-abelian groups found in the BPS unorientifold setting
\cite{bpstor}.  By arguments similar to those in
\cite{pw}, the apparent discrepancy between heterotic symmetry enhancement at
special values of the  radii and the lack of a similar phenomenon in the type I
description is to be interpreted as requiring D-brane states to become massless,
an effect that is beyond the reach of type I perturbation theory.

Precision tests of the duality between heterotic and type I
compactifications with reduced rank would require a detailed matching of
the spectrum of BPS\footnote{Here, BPS stands for
Bogomolny-Prasad-Sommerfield.} states or equivalently of the global
identifications in the moduli spaces of the compactifications. Indeed, although
supersymmetry alone does not fix uniquely  the massless spectrum in these
compactifications, it is still powerful enough to fix completely the low-energy
effective lagrangian once the spectrum is fixed. In particular the moduli space
parameterized by the scalars in the
$N_V$ vector multiplets and the scalar in the supergravity multiplet is
\cite{salsez}
\be  {O(d, N_V) \over O(d)\times O(N_V)}\times R^+
\label{modspace}
\ee  up to global identifications. In $D=4$  the factor $R^+$ is extended to
$SL(2,R)/U(1)$ after the complexification of the scalar in the ${\cal N}=4$
supergravity multiplet. For the component with maximal
rank $N_V = 16+d$ the heterotic string derivation of (\ref{modspace})
has been given long time ago \cite{nsw}. For the components with reduced
rank, that admit both a geometric interpretation and a candidate BPS
unorientifold dual, one can follow the arguments in \cite{cp}. In the type I
setting a discussion of the truly marginal deformations and of some of the global
identifications can be found in \cite{bpstor}. In Section 3 we expand on
that discussion and give an algebraic argument along the lines of \cite{dff} in
favour of the type I interpretation of (\ref{modspace}). For the time being, let
us notice that, contrary to the heterotic case, the scalar in the supergravity
multiplet is not simply the string dilaton but rather a combination of the
dilaton and the scalar parameterizing the volume of the compactification torus
\cite{chiral}. It is thus difficult to sort out the dilaton from the other
moduli. As a result, T-duality becomes a non-perturbative symmetry as expected
from D-brane considerations \cite{cjp}.

Below $D=6$, type II compactifications on $K3\times T^{d-4}$ give rise to
another class of superstring vacua with 16 supercharges. These models are
known as type II (2,2) models since in $D=4$ the ${\cal N}=4$ supersymmetry
charges are
evenly contributed from left and right movers \cite{sv,kk,vwdual}. In the
absence of any truncation, the gauge group is generically $U(1)^{2d+16}$. 
Using the well-established type IIA - heterotic duality in $D=6$ \cite{send,
senk} and mirror symmetry,
\ie T-duality, of the type II theories on $T^2$, one deduces that type IIA, type
IIB and heterotic strings are related by S-T-U triality \cite{send, asp}. As a
result the non-perturbative S-duality is mapped to perturbative T-duality. 
Type II duals of some CHL models can be obtained after orbifold
projections with a trivial action on the supercharges \cite{sv,ss,cl}.  
The detailed analyses \cite{asp,cl} show that there can be several models
with the same $N_V$ that, however, are expected to display 
different global identifications. Aside from the two models with $N_V=10$
in $D=4$ \cite{asp}, all the other instances have $N_V \leq d$. For the
models with $N_V<d$ it is seems difficult to
find a {\it perturbative} type I description. The components with $N_V=d$ can be
given a rationale in terms of type I superstrings without open strings
\cite{gepner, dp}. In these components the CP group is completely absent since
the O-planes that correspond to {\it unconventional}
$\Omega$-projections do not carry R$\times$R charge. The number of vector
multiplets that accompany the $d$ graviphotons is only
$d$ in this case. They correspond to the combinations $G_{\mu
i}-\widetilde B_{\mu i}$ orthogonal to the graviphotons. To the best of my
knowledge there is no way to reduce the rank of the type I gauge
group any further and to find perturbative type I duals of \eg the models with
$N_V=4$ in $D=4$ \cite{sv,cl,asp}. This should not necessarily sound as a
breakdown of heterotic - type I duality, but rather as a motivation for
exploring further these components of the moduli spaces.

In the type IIB superstring, the action of the world-sheet
parity projection $\Omega$  and the right-moving fermionic parity $(-)^{F_R}$
are conjugate to one another, 
$\Omega = S^{-1} (-)^{F_R} S $,
via an S-duality transformation\footnote{It would be more approriate to 
term this $S$ a U-duality
transformation since it involves a mixing between the NS$\times$NS dilaton
$\phi$ with the R$\times$R dilaton $\chi$.}:
\be
S: \qquad \lambda \rightarrow -1/\lambda \qquad ,
\label{sduality}
\ee
that inverts the (complexified) type IIB dilaton $\lambda = \chi + i
e^{-\phi}$. Although in $D=10$
self-duality of the type IIB superstring is not sufficient to prove
that the two orbifolds of the type IIB with respect to 
$\Omega$ and $(-)^{F_R}$ are dual to one another -- in fact one gets the type I
superstring and the type IIA superstring respectively \cite{send} --
an adiabatic argument \cite{send,vwdual,sv} suggests that one may recover
equivalence after toroidal compactifications when the action on the fields
is accompanied by a non trivial action on the geometry, \eg a shift $\sigma_V$
of order two in the compactified directions.  In Section 4, we will show that the
resulting dual pairs consist in type I models without open strings and type
II (4,0) models without D-branes. The latter correspond to asymmetric orbifolds
that break all the supersymmetries in the right-moving sectors. Models of this
kind have been considered both from the viewpoint of the fermionic construction
\cite{abk, fk, dkv} and from the viewpoint of the {\it freely-acting orbifolds}
\cite{kk}. At generic points of the moduli space the gauge group is
$U(1)^{d+d}$ that, at free fermionic points, gets perturbatively enhanced
to
$SU(2)^d\times U(1)^{d}$ or to another group whose structure constants satisfy
the constraints for the existence of a cubic world-sheet supercurrent
\cite{dkv,lls,abk}. Notice that these enhanced symmetry points typically
correspond to radii which are {\it half} the standard self-dual value
$R=\sqrt{\ap}$. This modified Halpern-Frenkel-Kac (HFK) mechanism is consistent
with the two following observations. First, the current algebra on the
world-sheet that leads to the enhanced symmetry is realized at higher level, \eg
$k=2$ in the simplest instance of $SU(2)^d$. Second, the T-duality group of
asymmetric orbifolds is different from the T-duality group of standard toroidal
or symmetric orbifold compactifications and the points of enhanced symmetry are
the self-dual points of the modified T-duality group. Notice that, rather
surprisingly, D-branes in type II (4,0) models do not carry R$\times$R charge. 
Indeed, 
the process of breaking all the supersymmetries in the right-moving sector all
the R$\times$R states, together with their NS$\times$R superpartners, become
massive. A nice feature, however, is that the twisted sectors of these orbifolds
do not give rise to massless particles in the decompactification limit, so that
these vacua regain the {\it rigidity} properties of their parent type II (4,4)
superstrings 
\cite{kk}. Using U-duality invariance in $D=6$, some type II (4,0) models have
been argued to be dual to type II (2,2) models \cite{sv,kk} and can thus be
connected to heterotic CHL models \cite{ss,asp,cl}.

In view of the mounting wave of interest in the putative M-theory
\cite{mt}, let us briefly comment on the relation of the above vacua to
M-theory compactifications. According to common wisdom, M-theory is the strong
coupling limit of the type IIA superstring in $D=10$ \cite{witdyn} or
alternatively of the
$E(8)\times E(8)$ heterotic string in $D=10$ \cite{horwit}. Both
equivalences follow from the identification of the vacuum expectation value of
the dilaton with the length of an extra dimension. Although the former
equivalence ascribes both the massless spectrum and low-energy
effective lagrangian from a dimensional reduction {\it \'a la} Kaluza-Klein
(KK), the latter rests on further assumptions such as residual
supersymmetry,  anomaly cancellation {\it and} a consistent string
interpretation. For instance, relaxing the last assumption, the
${\cal N}=(1,0)$ supergravity theory with gauge group
$U(1)^{248}\times U(1)^{248}$ \cite{gsw} could not be excluded. Still, it is
remarkable how simply assuming the existence of a consistent 11D theory has
provided so many non-perturbative connections in the web of superstring vacua
\cite{witdyn}. In fact, it has been shown that the type I vacua without open
strings are equivalent to M-theory compactifications on non-orientable manifolds
without boundaries \cite{dp}. In particular, the type I superstring without open
strings in $D=9$ is equivalent to M-theory on a Klein bottle. By the same
token, M-theory compactification on the M\"obius
strip is expected to describe the strong-coupling limit of the CHL strings
in $D=8$ \cite{dp,park} that are related to the BPS
models with $b=2$. Another class of compactifications of M-theory that are
supposed to preserve 16 supercharges and have received some attention in view
of heterotic - type IIA duality in
$D=6$ \cite{witdyn, senk, send} consists of compactifications on $K3\times S^1$
and orbifolds thereof
\cite{ss,cl,sv,kr}. Prior to any truncation one simply gets type IIA on $K3$
with 20  ${\cal N}=(1,1)$ vector multiplets. Modding out by a symmetry that
preserves the holomorphic 2-form on $K3$ and acts as a shift on the extra
circle gives rise to other ${\cal N}=(1,1)$ supersymmetric vacua with a gauge
group of reduced rank. The complete classification \cite{ss,cl}
include four possibilities with $N_V = 20, 12, 8, 4$ that can be accounted
for by the BPS unorientifolds with
$b=0,2,4$ and the type I vacua without open strings. Once again we remark that
the components of the moduli space of vacua with 
$N_V=2$ \cite{ss,cl} do not seem to have an obvious {\it perturbative} type I
interpretation.

Another interesting class of superstring vacuum configurations goes under the
name of F-theory \cite{ft}. Almost by definition, F-theory on a manifold
${\cal F}$ that admits an elliptic fibration, \ie looks locally like
${\cal F} = {\cal B}\times T^2$, is the compactification of the type IIB
superstring on the manifold ${\cal B}$ with 24 7-branes. The complexified
dilaton $\lambda$ is identified with the complex modulus
of the elliptic fiber $T^2$ and its variation over the basis
${\cal B}$ is governed by the arrangement of 7-branes. It is a remarkable fact
that under two T-dualities the 16 dynamical D9-branes, present in the
type I compactification on $T^2$, are mapped to as many D7-branes while the
O7-planes may be regarded as bound states of two 7-branes each
\cite{senf}.  The arrangement of 7-branes can be chosen so that 
$\lambda$ is constant and small and the resulting configuration is T-dual to
a perturbative type I description \cite{senf,dm}. F-theory vacua may thus be used
as non-perturbative definitions of the dynamics of D-branes and O-planes in type
I vacua
\cite{senf,dm,sengp}. Moreover, the conjectured duality between the heterotic
and type I superstrings allows one to establish dualities between F-theory and
heterotic compactifications. In particular, F-theory on an elliptic $K3$ in
the orbifold limit $K3\approx T^4/Z_2$ may be related to the heterotic string
on $T^2$
\cite{senf}. Indeed the moduli spaces of elliptic $K3$ surfaces is
$O(2,18)/O(2) \times O(18)$  and, up to global identifications, coincides with
the trivial component of the moduli space of heterotic string
compactifications on $T^2$ \cite{nsw}. By fiberwise application of the above
duality \cite{vwdual}, one can identify F-theory compactifications on elliptic
Calabi-Yau threefolds and fourfolds with  heterotic and type I compactifications
on Calabi-Yau twofolds and threefolds in the presence of penta-branes
\cite{ft,blum,gepner,sengp}. It is beyond the purposes of the present
investigation on vacua with 16 supercharges to pursue this viewpoint, that is the
subject of active research in view of the possibility of extracting
some non-perturbative information on vacua with
${\cal N}=2$ (8 supercharges) and ${\cal N}=1$ supersymmetry (4
supercharges) in $D=4$. There are however threefolds and fourfolds of reduced
holonomy
\cite{asp,park,blum}. In such cases one gets enhanced supersymmetry both in
$D=6$ and in $D=4$. According to the classification in
\cite{asp}, in $D=4$ one finds $N_V=22, 14,
10, 6, 4$. Aside from the last case, the values of $N_V$ precisely match with 
those found in BPS unorientifolds \cite{bpstor} and in type I theories without
open strings \cite{dp,gepner}.
 
One last remark on the role of a quantized NS$\times$NS antisymmetric tensor
background in type I compactifications concerns the case ${\cal N}=(1,0)$ in
$D=6$. The first model of this kind \cite{bssys} had a CP group $Sp(8)^4$
that, though {\it smaller} than expected from a na\"ive analogy with heterotic
vacua with perturbative embedding of the spin connection in the gauge group, is
not a subgroup of $SO(32)$. It is by now clear that the {\it oblique} CP
symmetry enhancement is due to the presence of D5-branes \cite{cjp}. What
may still seem  puzzling is the effective reduction by a factor of two in the
number of both D5-branes and D9-branes! In order to solve this puzzle, 
one has to
recall that the model in question descends from the type IIB compactification on
a
$Z_2$-orbifold of $T^4$ at the
$SO(8)$ enhanced symmetry point \cite{bssys}. Since the rank of the
NS$\times$NS antisymmetric tensor that correspond to the $SO(8)$ current
algebra at level
$k=1$ is $b=2$, one could expect a reduction by a factor of two of the overall
CP multiplicities, while the symplectic nature of the CP group does the rest. A
similar analysis can be performed for the models in \cite{bssys,gepner,
bstw} that corresponds to rational points in the moduli space of 
orbifold compactifications \cite{cjp,dp}. In particular for models with one
tensor multiplet, that admit perturbative heterotic duals, it has been 
shown
\cite{sese} that a quantized NS$\times$NS antisymmetric tensor corresponds to a
compactification with a generalized second Stiefel-Whitney
class\footnote{For a $Spin(32)/Z_2$ 
vacuum gauge bundle ${\cal V}$, $\omega_2(\cal V)$ represents the obstruction to
defining a vector structure, \ie a consistent parallel transport for fields in
the representations of $SO(32)$ that belong to the vector conjugacy class 
\cite{tutti}.}
$\omega_2({\cal V})$ that satisfies $\omega_2({\cal V}) = 2B$. 
Since $\omega_2({\cal V})$ belongs to
$H^2({\cal S},Z_2)$, where ${\cal S}$ is the $K3$ surface under consideration,
and $B$ is defined modulo shifts in $H^2({\cal S},Z)$, one has three
inequivalent choices \cite{sese}. Upon performing T-duality on the two-cycle
with non-vanishing $B$-flux one ends up with an F-theory compactification with
vanishing $B$ and a mirror $K3$ surface $\tilde{\cal S}$ \cite{sese}.

Clearly, in the long run, one would like to address the issues raised by the
presence of a quantized NS$\times$NS antisymmetric tensor in type I
compactifications with ${\cal N}=1$ supersymmetry in $D=4$. A preliminary
analysis has been performed for the type I descendandants of the type IIB
superstring on the $Z$-orbifold \cite{chiral,kak}. The 6D cases
should be taken as a guide to explore further connections 
in $D=4$ between BPS unorientifolds and other consistent superstring vacua.
Relation to M-theory compactifications and F-theory vacua
\cite{senf,blum,twelve} may help understanding geometric features, \eg the
moduli space of vacua, that sometimes look obscure from a superstring
perspective. 

The plan of the paper is as follows. Section 2 is a primer on rational
unorientifolds that motivated the discovery of the BPS unorientifolds. 
In Section
3 we discuss generalized toroidal compactifications of the type I superstrings
and show how a quantized NS$\times$NS antisymmetric tensor may be interpreted in
terms of non-commuting open-string Wilson lines. We also discuss the local
structure of the moduli space of BPS unorientifolds and their global
identifications. Finally, we derive the one-loop partition function of the CHL
model with
$E(8)$ current algebra at level $k=2$ for the sake of comparison with the
BPS unorientifolds in $D=8$. In Section 4 we discuss type II superstring vacua
with 16 supercharges and argue that some type II (4,0) models without
D-branes are dual to unconventional type I vacuum configurations without open
strings. Finally, Section 5 contains some speculations and our conclusions.
 
\section{Quantized $B$ from Rational Un-orientifolds}
 
The building blocks of perturbative closed-string theories are conformal field
theories on closed orientable Riemann surfaces \cite{bpzfms,gep}. The building
blocks of perturbative open-string theories are conformal field
theories on closed, open and/or unorientable Riemann surfaces\footnote{Early
calculations of multi-loop open-string scattering amplitude in
terms of Green functions on surfaces with boundaries date back to the work of
Alessandrini and Amati \cite{aa}.}. The discovery by Green and Schwarz of
infinity and anomaly cancellations in the type I superstring with gauge group
$SO(32)$ \cite{gs}, that triggered an enormous interest in the field and led
to the discovery of the heterotic string \cite{ghmr}, also motivated the
discovery of the $SO(8192)$ bosonic string \cite{ms}. The proposal of
interpreting open-string theories as descendants of left-right symmetric
closed-string theories \cite{as} was developed in \cite{bsrmg,ps} and brought to
a consistent systematization in \cite{bssys,bstw}. For rational models, the
crucial issue of CP symmetry breaking was achieved borrowing some interesting
results of Cardy's on boundary effects in two-dimensional critical models
\cite{jc}.  

The starting point for the construction of a rational {\it un-orientifold} is a
left-right symmetric rational conformal field theory (RCFT). Rationality is
related to the presence of a chiral algebra ${\cal C}$ of symmetries on the
world-sheet, \eg a current algebra, that extends the Virasoro algebra generated
by the modes $L_n$ of the energy-momentum tensor $T(z)$ and allows to encode the
spectrum of the theory in a {\it finite} number of characters    
\be 
\chi_h(q) = Tr_{_{{\cal H}_h}} q^{L_o-{c\over 24}} \qquad ,
\label{spectrum}
\ee 
As usual $c$ is the central charge of the Virasoro algebra, $q=\exp(2\pi i
\tau)$, with $\tau$ the one-loop modular parameter and
${{\cal H}_h}$ denotes the sector of the spectrum formed by the descendants
(with respect to the chiral algebra ${\cal C}$) of the primary field with
$L_o=h$. The characters $\chi_h(q)$ provide a (unitary) representation of the
modular group $SL(2,Z)$ generated by the transformations:
\be
S: \qquad \chi_h(-1/\tau) = S_{hk}\chi_k(\tau)
\label{modstrans}
\ee
and
\be
T: \qquad \chi_h(\tau + 1) = e^{2\pi i \left(h-{c/24}\right)} \chi_h(\tau)
\label{modttrans}
\ee
and enter the torus partition function in a modular invariant way
\be   
{\cal T} = \sum_{h\hbar} N_{h\bar h} \chi_h (q) \chi_{\bar h}({\bar q})
\label{torus}  
\ee 
where $N_{h\bar h}$ are integers that satisfy $N_{oo}=1$, where $o$ labels
the identity primary field with $L_o=0$. 

In order to construct an open-string descendant of an
oriented closed-string model based on a RCFT, that is invariant under 
left-right interchange, 
\ie
$N_{h,\bar h}=N_{\bar h h}$, one starts by dividing {\cal T}
by a factor of two. The $\Omega$-projection introduces O-planes \cite{cjp} that
are accounted for by the Klein-bottle amplitude 
\be  
{\cal K} = {1\over 2}\sum_{h} N_{hh} \s_h \chi_h(q\bar q)   
\label{klein}
\ee  
that completes the {\it untwisted sector} of the un-orientifold \cite{as}:
\be
{\cal Z}_u = {1\over2} Tr_{{\cal H}_c} (1+\Omega) q^{L_o-{c\over 24}}
\bar q^{\bar L_o-{c\over 24}} = {\cal T} + {\cal K} \quad ,
\ee
where as indicated the trace is taken over the closed-string states. 
The signs $\s_h$ in the $\Omega$-projection \ref{klein} are restricted by the
{\it crosscap constraint}
\cite{fpss} that \eg requires $\s_i\s_j=\s_k$ if the
fusion-rule coefficient $N_{ij}^k$, given by the Verlinde formula
\cite{ev}, is non-vanishing. 

To the resulting unoriented closed-string spectrum one can -- and in many cases
must -- add the unoriented open-string spectrum. Observing that
$q\bar q= q_o^2$, with $q_o=\exp(-2\pi\tau_2)$, the most general
parameterization of the {\it twisted sector} of the un-orientifold \cite{as} is 
\be
{\cal Z}_t = {1\over2} Tr_{{\cal H}_o} (1+\Omega) q_o^{L_o-{c\over 24}}
 = {\cal A} + {\cal M} \quad ,
\label{open}
\ee
where as indicated the trace is taken over the open-string states. 
More explicitly, (\ref{open}) involves the annulus partition function: 
\be  {\cal A} = {1\over 2} \sum_{h,a,b} A_{ab}^h n^a n^b \chi_h(\sqrt{q_o})
\quad , 
\label{annulus}
\ee
where $n^a$ are the CP multiplicities and $A_{ab}^h$ are integer coefficients,  
and the M\"obius strip $\Omega$-projection 
\be  {\cal M} = {1\over 2} \sum_{h,a} M_{aa}^h n^a
\widehat\chi_h(e^{i\pi}\sqrt{q_o})  
\label{moebius}
\ee  
where $M_{aa}^h= A_{aa}^h (mod2)$ and $\widehat\chi_h$ form a proper basis of
{\it hatted} characters
\cite{bssys} 
\be
\widehat\chi_h (i\tau_2 + 1/2) = e^{-i\pi (h-c/24)} \chi_h (i\tau_2 + 1/2)
\quad ,
\ee  
real functions of
$\tau_2$, thanks to the overall phase-shift. Although in general the CP factors
$n^a$ and the sectors of spectrum
${\cal H}_h$ cannot be put in one-to-one correspondence, for the
charge-conjugation modular invariant ($N_{h,\bar h}=C_{h,\bar h}= \delta_{h,\bar
h^c}$) one is allowed to associate to each sector ${\cal H}_i$ a CP factor
$n^i$ and let $A_{ij}^k=N_{ij}^k$ in (\ref{annulus}) or an
automorphism thereof
\cite{bssys,bstw}. Sewing of surfaces with holes and crosscaps implies some
consistency conditions on the above parameterization. Most notably, the
{\it completeness conditions} $A_i^a{}_c A_j^c{}_b=N_{ij}{}^kA_k^a{}_b$
\cite{fpss}. 

After switching to the transverse closed-string channel via a modular
S-transformation (\ref{modstrans}) the Klein-bottle amplitude (\ref{klein}) gives
the crosscap-to-crosscap amplitude 
\be 
\widetilde{\cal K} = \sum_h (\Gamma_h)^2 \chi_h  
\label{crosscap}
\ee  
that, up to some sign ambiguity, allows one to extract the crosscap reflection
coefficients $\Gamma_h$, determining the coupling of the $h$-sector of the
closed-string spectrum to O-planes. Similarly, via a modular
S-transformation (\ref{modstrans}), the annulus partition function
\ref{annulus} gives the boundary-to-boundary amplitude  
\be 
\widetilde{\cal A} = \sum_h (B^h)^2\chi_h  = \sum_{h,a} (B_{ha} n^a)^2 \chi_h 
\quad ,
\label{boundary}
\ee
where $B_h$ are the boundary reflection coefficients that determine the
coupling of the $h$-sector of the closed-string spectrum to 
D-branes.
 Given the amplitudes (\ref{crosscap}) and (\ref{boundary}), the consistency of
the construction requires that the boundary-to-crosscap amplitude be of the
form 
\be 
\widetilde{\cal M} = \sum_{h,a} \Gamma_h (B_{ha} n^a) \widehat\chi_h 
\label{btc}
\ee   
The modular transformation between loop and tree channel of the
M\"obius strip amplitude is induced by  $P = T^{1/2} S T^2 S T^{1/2}$ that acts
on {\it hatted} characters  and satisfies  $P^2 = C$ \cite{bssys}. The
required form of (\ref{btc}) put severe constraints on the choices of integer
coefficients
$A_{ab}^h$ that appear in (\ref{annulus}) and of the various signs 
in the $\Omega$-projection both in the {\it untwisted} and in the {\it twisted}
sectors. For the charge-conjugation modular invariant, these are seen to be all
fulfilled by the choice $A_{ij}^k=N_{ij}^k$ \cite{bssys,fpss}.

In order for a rational un-orientifold  
to be interpreted as a consistent open-string vacuum configuration, one has to
impose the proper connection between spin and statistics \cite{bsrmg} and
cancellation of the tadpoles of the unphysical massless states
\cite{pc,bssys}. The latter require $\Gamma_h + B_{ha} n^a = 0$, for all
the massless states that have been eliminated 
by the $\Omega$-projection in the direct closed-string channel. 
In imposing these conditions it is crucial to fix the relative 
normalization of ${\cal K}$, ${\cal A}$ and ${\cal M}$ by expressing them
in terms of the modular parameter of the common double-cover \cite{ms}.
In modern language, this amounts to R$\times$R charge neutrality of the
relevant configuration of D-branes and O-planes
\cite{cjp}. For non-supersymmetric configurations, tadpole cancellation for
other massless physical fields, such as the dilaton,  can be unambiguously
imposed as a requirement for vacuum stability
\cite{ms}. The simplest {\it essentially} rational closed-string theory one can
un-orientifold is the type IIB superstring in
$D=10$. The result is the type I superstring with gauge group
$SO(32)$ \cite{as}. In $D=10$ there are two more left-right symmetric
theories: the tachyonic models proposed long time ago
\cite{swtach}. Their open-string descendants \cite{bssys} play a crucial role
in some proposed string dualities without supersymmetry \cite{bd} and hopefully
\cite{mbik} may provide a rationale for the largely unexplored ${\cal N}=(1,0)$
10D supergravity with gauge group $U(1)^{496}$ \cite{gsw}. 

In order to display the subtlety that allows for the existence of the BPS
unorientifolds one has to consider generalized $\Omega$-projections of type II
compactifications \cite{bssys,bstw} that are compatible with target-space Lorentz
symmetry and with the  diagonal part of the internal symmetries\footnote{One may
also envisage the possibility of un-orientifolds that break some of the internal
{\it accidental} symmetries. Some instances are discussed in \cite{bstw}.}. In
particular, the free fermionic constructions
\cite{abk} or the covariant bosonic lattices \cite{lls} allow for simple and
elegant rational compactifications. The intrinsic consistency of type I
descendants of certain left-right symmetric type II models forces one to allow
for the introduction of a non-vanishing but quantized NS$\times$NS antisymmetric
tensor background. For instance, closed-string models that involve a
simply-laced current algebra of rank $r$ at level $k=1$  
correspond to propagation on an $r$-dimensional 
torus with constant internal metric $G_{ij}$ identified with the
Cartan matrix
$C_{ij}$ of the underlying Lie algebra and, more importantly for our goals,
with the NS$\times$NS antisymmetric tensor satisfying
$B_{ij}=C_{ij}$ for
$i>j$ and $B_{ij}=-C_{ij}$ for $i<j$ \cite{lls}. The left-right symmetry of the
theory that becomes apparent when the torus partition function of the parent
closed-string theory is written as in (\ref{torus}) suggests the possibility of
introducing a quantized NS$\times$NS antisymmetric tensor at generic points of
the moduli spaces of toroidal or orbifold compactifications
of their open-string descendants\cite{bpstor,gepner}.

\section{Toroidal Compactifications Revisited}

In order to analyze generalized toroidal compactifications of the type I
superstring
\cite{bpstor} let us start with a discussion of the conditions for left-right
symmetry of the parent type IIB compactifications. Un-orientifolds
of type IIA models simply follow from T-duality
\cite{cjp,hor,pw}. Since the spectrum of oscillator excitations is automatically
invariant under
$\Omega$, the only potential troubles come from the Narain lattice
$\Gamma_{(d,d)}$ of generalized momenta $(P_L, P_R)$. Left-right symmetry
implies that for any state
$(P_L, P_R)$ there exists a {\it specular} state $(P_R, P_L)$, \ie a state with 
$P_L^{\prime}= P_R$ and $P_R^{\prime}= P_L$. Given a generic value of the
metric $G_{ij}$, this turns out to be a constraint on the NS$\times$NS
antisymmetric tensor $B_{ij}$ \cite{bpstor}. Indeed using the standard
parameterization\footnote{In order to adhere to the recent literature
\cite{sese} we have changed the normalization of $B_{ij}$ by a factor of two
with respect to \cite{bpstor} and put $\ap=2$.} 
\be P^i_{L/R}(m,n) = G^{ij} (m_j + B_{jk} n^k) \pm {n^i \over 2}
\label{momenta}
\ee 
with integer $n^i$ and $m_j$ and $G^{ij}$ the inverse of $G_{ij}$, and
imposing $P_L(m,n) = P_R(m^\prime,n^\prime)$ for generic $G_{ij}$, one
immediately finds 
\be n^{i\prime} = - n^i  \qquad   m_j - m^\prime_j + 2 B_{jk} n^k = 0
\label{bquant}
\ee  
the second condition implies that $2B$ belongs to the integer cohomology of the
torus $T^d$ and determines $n^k$ in terms of $m_i$ and $m_i^\prime$. Given these
constraints and since only states with $P_L= P_R$, \ie Kaluza-Klein (KK)
momentum states with $n^i=0$, are fixed under $\Omega$,
one can check that the Klein-bottle contribution ${\cal K}$ does not depend on
$B_{jk}$. This should not sound unexpected, given the inconsistency of
pulling $B$ back to an unorientable surface.  

The perturbative unoriented closed-string spectrum is trivially
invariant under constant {\it continuous} PQ shifts of the
internal R$\times$R antisymmetric tensor 
\be
\widetilde B_{ij} \rightarrow \widetilde B_{ij} + \Delta_{ij}
\label{pq}
\ee  and less trivially invariant under constant {\it discrete} $GL(d,Z)$
transformations of the metric {\it and} the {\it quantized} NS$\times$NS
antisymmetric tensor 
\be G_{ij}+ B_{kl} \rightarrow (M^{-1})_i{}^k (G_{kl}+ B_{kl}) M^l{}_j 
\label{gltrans}
\ee  The latter allow to skew-diagonalize $B_{ij}$, \ie bring it to a form
$\widehat B_{ij}$ with only non-vanishing components equal to $0$ or $\pm
1/2$ in the diagonal two-by-two blocks. For later use, notice that $rank(B)
$ is invariant under $GL(d;Z)$ transformations. 

The open-string spectrum presents some amusing features \cite{bpstor}. First
of all, thanks to the possibility of adding Wilson lines $A_i^a$ in the Cartan
subalgebra  of $SO(32)$, the KK momenta are shifted by an amount $q_a A_i^a$ as
expected.  Rather unexpectedly, however, they can suffer a further shift due the
presence of a quantized $B_{ij}$. In order to deduce this shift one has to start
from the transverse closed-string channel where Wilson lines become phases in
the boundary reflection coefficients. The presence of boundaries and/or
crosscaps requires the constraint $P_L = - P_R$ on the closed-string states
flowing along the tube. To this end, one has to introduce exactly
$b=rank(\widehat B)$ $Z_2$-projections 
\be  {\cal P}_i = {1\over 2} \left(1 + (-)^{(2\hat B n)_i} \right) \qquad .
\label{project}
\ee  
After imposing the cancellation of the unphysical tadpole in the
R$\times$R sector, the CP multiplicity is reduced by a factor $2^{b/2}$, as
stated in the introduction and found some time ago \cite{bpstor}. For
instance, in $D=8$, one can take advantage of the {\it discrete} choices of $B$
and the {\it continuous} Wilson lines to connect the BPS unorientifolds with
$G_{CP}=SO(16)$ to the ones with 
$G_{CP}=Sp(16)$ passing through points where the CP symmetry is broken to
$U(1)^8$ or is partially enhanced to $U(8)$ \cite{bpstor}.

The existence of non-perturbative D-brane states \cite{cjp} breaks the {\it
continuous} PQ shifts of the R$\times$R  antisymmetric tensor to  {\it
discrete} ones. In type II compactifications with maximal supersymmetry the
(pseudo)scalars from the R$\times$R sector transform according to a spinorial
representation of the
$O(d,d)$ T-duality subgroup of the full $E_{(d+1)}(d+1)$ U-duality group
\cite{dff}. Under the $GL(d)$ subgroup of perturbative type I dualities the
spinorial representation decomposes into a sum of rank $p$ antisymmetric
tensors. In particular $\widetilde B_{ij}$ transforms as a two-index
antisymmetric tensor as it should. Similarly  $A_i^a$ transforms according to
the $({\bf d}, {\bf n})$ representation of $GL(d) \times O(n)$, where $n$
is the rank of the surviving CP
group. Discrete shifts of $A_i^a$ can be compensated by discrete shifts of
the KK momenta $m_i$ and are thus symmetries of the spectrum \cite{bpstor}.

By the same line of arguments as in \cite{dff}, one can determine the moduli
space of the compactification, up to global identifications. Indeed the exact
perturbative symmetries of the type I unorientifolds, \ie the decoupling at
zero-momentum of the moduli fields from all perturbative
scattering amplitudes \cite{bpstor}, allow one to put
$G_{ij}$, $\widetilde B_{ij}$, 
$A_i^a$ and the dilaton $\Phi$ in one-to-one correspondence with the generators
of a solvable Lie algebra, that upon exponentiation produces the moduli space
(\ref{modspace}).

The global identifications that form the smallest group that combines
the discrete shifts of
$\widetilde B_{ij}$ and $A_i^a$ with the $GL(d,Z)\times O(n;Z)$ transformations
is the expected U-duality group
$O(d,d+n;Z)$\footnote{This conclusion has been reached in collaboration
with Carlo Angelantonj and Yassen Stanev.}. 
Notice that the identification of the U-duality group relies on the assumption
of R$\times$R-charge integrality for non-perturbative D-brane states.
This assumption allows for discrete but otherwise arbitrary shifts of $\widetilde
B_{ij}$. Moreover the $O(d,d;Z)$ subgroup is
not the perturbative T-duality group of the parent type II superstring. It is
instead the subgroup of the full U-duality group $E_{(d+1)}(d+1;Z)$ that
involves mixing $G_{ij}/\sqrt[d]{det(G_{ij})}$ with
$\widetilde B_{ij}$ and a certain combination of the dilaton $\Phi$ and
$det(G_{ij})$. The scalar $\Phi_H$ parameterizing the $R^+$ factor of the type
I moduli space follows from an educated use of heterotic - type I duality in
$D=10-d$ dimensions and coincides with the combination of
$\Phi$ and $det(G_{ij})$ that gives the heterotic dilaton \cite{chiral}:   
\be 
\Phi_H = {6-D \over 4} \Phi_I - {D-2 \over 16} ln(det G_I) \qquad .
\label{relax}
\ee

One can interpret the observed rank reduction of the CP group in the BPS
un-orientifolds as associated to non-trivial holonomies for unconventional 
type I superstring propagation on $T^d$ with quantized NS$\times$NS
antisymmetric tensor much in the same way as in unconventional type I
compactifications to
$D=6$ on $K3$ \cite{sese}. Indeed one can define the generalized second
Stiefel-Whitney class of the vacuum gauge bundle ${\cal V}$, $\omega_2({\cal
V})$, via the relation
\be W(\gamma_1) W(\gamma_2)= 
\exp\left({i \pi \int_\Sigma \omega_2({\cal V})}\right) W(\gamma_2) W(\gamma_1)
\qquad ,
\label{holonomy}
\ee 
where $W(\gamma_1)$ and $W(\gamma_2)$ are Wilson loops and, for the toroidal case
under consideration, the two-cycle $\Sigma$ is the product of the one-cycles
$\gamma_1$ and
$\gamma_2$. Recall that for $Spin(32)/Z_2$, the expected non-perturbative gauge
group of the type I superstrings,
$\omega_2({\cal V})$ represents the obstruction to defining a vector structure
\cite{tutti}. The relation $B={1\over 2}\omega_2$, up to a shift in integer
cohomology, follows from duality between open- and closed-string viewpoints in
the BPS unorientifolds. Indeed the phase-shift a closed-string experiences in the
presence of a quantized flux of $B$
\be
\exp\left({2i \pi \int_\Sigma B}\right)
\label{flux}
\ee 
must exactly match the phase in (\ref{holonomy}) \cite{sese}. In a recent
paper \cite{lmst} the geometry of the CHL strings has been interpreted in
terms of non-commuting Wilson lines. Another independent analysis can be found in
\cite{kak}. Still, the type I description in terms of a quantized NS$\times$NS
antisymmetric background and the inextricable link between open and closed
unoriented strings seem by far to provide the most economical explaination. It
may also have far reaching consequences in {\it phenomenologically viable}
compactifications, since
${\cal N}=1$ supersymmetric heterotic models in $D=4$ with adjoint Higgses
often require higher level current algebras on the world-sheet. Higher level
current algebras and their by-product in terms of adjoint Higgses should
require a non-vanishing $\omega_2({\cal V})$ in Calabi-Yau compactifications as
well.  Their type I duals are thus
expected to correspond to ${\cal N}=1$ supersymmetric unconventional type I
compactifications to
$D=4$ such as those briefly discussed in \cite{chiral,kak}.
 
An alternative interpretation of the above phenomenon has been given in terms
of D-branes for the case $D=8$ \cite{dp, park}. For completeness and for the
sake of comparison let us briefly run the argument backwards. Performing a
T-duality transformation along one of the two one-cycles of $T^2$ not only
turns the type IB theory\footnote{We adopt the terminology recently suggested
by Clifford Johnson.} into the type IA theory but also trades the NS$\times$NS
antisymmetric tensor for an off-diagonal component of the
metric. The latter is to be
associated to an effective reduction by a factor of two in the volume of the
elementary cell defining the torus. The same reduction by a factor two
results in the number of fixed {\it lines} of the new $\Omega$-projection and
the neutral configuration then requires only half the original number of
D8-branes and gives rise to a CP group $SO(16)$ \cite{park,kak}. By the same
line of arguments as after eq.(\ref{project}) one can continuously pass to
$Sp(16)$ \cite{kak}.

For a direct comparison with the heterotic string, let us consider  the
simplest CHL model that corresponds to breaking $E(8)\times E(8)$ at level
$k=1$ to the diagonal $E(8)$ subgroup at level $k=2$ in $D=8$. Although the
analysis of the massless spectrum has been performed in \cite{cl},  we derive
the genus-one partition function in order to streamline the striking
similarity with the un-orientifold construction
\cite{as}. Denoting by
$\tau$ the modulus of the worldsheet torus and neglecting the invariant measure
$|d\tau|^2/ \tau_2^2$, the partition function for the conventional toroidal
compactification to $D=8$ \cite{nsw} reads  
\be Z_{T} = (\bar V_8 - \bar S_8) 
\left(\sum_{W \in \Lambda} {q^{W^2/ 2} \over \eta (q)^8 } \right)^2 
\left(\sum_{P=(p,\bar p) \in \Gamma} {q^{p^2/2} \bar q^{\bar p^2/2}\over
\tau_2^3|\eta (q)^8|^2 } \right) \qquad ,
\label{hettor}
\ee  where $\Lambda$ is the root lattice of $E(8)\times E(8)$ and $\Gamma$ is the
compactification lattice of generalized momenta in (\ref{momenta}). 
For compactness, we have also introduced the four characters
$\{V_8, O_8,S_8,C_8\}$ of the $SO(8)$ current algebra at level one, that encode 
the contribution of the world-sheet fermions and can
be easily expressed in terms of $\theta$-functions \cite{bssys}. 

 An orbifold projection that preserve all the 16 supercharges corresponds to
an order-two shift $\s_V$ of the coordinates of $T^2$ along a vector
$V=(v,\bar v)$, accompanied by the exchange $\widetilde\Omega$ of the two
$E(8)$ factors. Only states with the same $E(8)$ weight in (\ref{hettor})
are fixed under $\widetilde\Omega$, so that the untwisted sector
reads:
\be
Z_{++} = {1\over 2} Z_{T}
\label{hetpp}
\ee
and the $\widetilde\Omega$-projection yields
\be  Z_{+-} = {1\over 2}(\bar V_8 - \bar S_8) 
\left( \sum_{W \in \Lambda} {(q^2)^{W^2/2} \over \eta (q^2)^8 } \right) 
\left(\sum_{P=(p,\bar p) \in \Gamma} e^{2\pi i V\cdot P} {q^{p^2/2} \bar
q^{\bar p^2/2}\over
\tau_2^3|\eta (q)^8|^2 } \right) \qquad .
\label{hetpm}
\ee 
Performing an modular S-transformation, \ie $\tau \rightarrow -1/\tau$,
on (\ref{hetpm}) and a Poisson resummation on $(p,\bar p)$ yields the twisted
sector:
\be  Z_{-+} = {1\over 2}(\bar V_8 - \bar S_8) 
\left( \sum_{W \in \Lambda} {(\sqrt{q})^{W^2/2} \over \eta (\sqrt{q})^8 }
\right) 
\left( \sum_{P=(p,\bar p) \in \Gamma}{q^{(p+v)^2/2}
\bar q^{(\bar p + \bar v)^2/2}\over
\tau_2^3|\eta (q)^8|^2 } \right)
\label{hetmp}
\ee Finally performing a modular T-transformation, \ie $\tau \rightarrow
\tau +1$, yields the projection of the twisted sector:
\be  Z_{--} = -{e^{i\pi V\cdot V}\over 2}(\bar V_8 - \bar S_8) 
\left( \sum_{W \in \Lambda} {e^{i\pi W\cdot W \over 2 }(\sqrt{q})^{W^2/2}
\over e^{-{i\pi\over 3}} \eta (e^{i\pi} \sqrt{q})^8 } \right)  
\left( \sum_{P=(p,\bar p) \in \Gamma} e^{2 i \pi V \cdot P} 
{q^{(p+v)^2/2}\bar q^{(\bar p + \bar v)^2/2}\over\tau_2^3|\eta (q)^8|^2}
\right) 
\label{hetmm}
\ee Notice the striking similarity of the $\widetilde\Omega$-projection with
the
$\Omega$-projection \cite{as,bssys} described in Section 2. Modular invariance
requires $V \cdot P$ to be half-integer and
$V\cdot V$ to be integer. Generically, the twisted sector does not contribute 
massless states while the untwisted sector contributes the supergravity
multiplet (including two graviphotons), the vector multiplets for the diagonal
$E(8)$ and two generically abelian vector multiplets. Adding Wilson lines
symmetrically in the two $E(8)$'s, \ie
$A_1=A_2=(1,(0)^7)$, allows one to break the diagonal $E(8)$ to $SO(16)$
\cite{cl}. By varying the Wilson lines one can continuously pass to $Sp(16)$ or
to $U(8)$ thus reproducing symmetry enhancements observed in BPS
un-orientifolds in $D=8$ \cite{bpstor}.

\section{Type II models with 16 supercharges }

There are two classes of type II string vacua with 16 supercharges. The first
class corresponds to type II (2,2) strings, that have half of the
supersymmetries carried by the left-movers and half by the right-movers, and
is available, as stated in the introduction, for $D\leq 6$. In fact models in
this class include compactifications on $K3\times T^{d-4}$ or freely
acting orbifolds thereof \cite{kk,sv}. The full perturbative spectrum is encoded
in the genus-one partition function, that may be easily derived in the
abelian orbifold limits of $K3$, following the standard procedure for symmetric
orbifolds, or at rational points where the compactification corresponds to a
tensor product of
${\cal N}=2$ superconformal minimal models, following the procedure devised by
Gepner \cite{gep}.  

The second class, that corresponds to the type II (4,0) strings, includes
asymmetric orbifolds of tori. In the notation introduced above, the one-loop
partition function for the toroidal compactification of the type IIB
superstrings reads 
\be  Z_{T} = (\bar V_8 - \bar S_8) ( V_8 - S_8)
\sum_{P=(p,\bar p) \in \Gamma} {q^{p^2/2} \bar q^{\bar p^2/2}\over
\tau_2^3|\eta (q)^8|^2 } 
\label{twopp}
\ee 
In order to be explicit but at the same time to keep the discussion as simple
as possible, let us restrict our attention to a $Z_2$-orbifold that acts as
$(-)^{F_R} \times \sigma_V$, where
$\sigma_V$ is a shift of order two in the lattice $\Gamma$, and thus preserves
all the 16 left-moving supercharges.  The $Z_2$-projection of the
untwisted sector yields:
\be
Z_{++} = {1\over 2} Z_{T}
\ee
and
\be Z_{+-} = {1\over 2}(\bar V_8 - \bar S_8) ( V_8 + S_8)
\sum_{P=(p,\bar p) \in \Gamma} e^{2\pi i V \cdot P} {q^{p^2/2} \bar
q^{\bar p^2/2}\over
\tau_2^3|\eta (q)^8|^2 } 
\label{twopm}
\ee Due to the change of sign in $S_8$, in
the NS$\times$R sector and, more interestingly, in the R$\times$R sector only
massive states with  $V \cdot P$ an odd integer survive the projection, so that
these type II (4,0) strings  are in fact type II superstrings without D-branes!
For the purpose of discussing gauge symmetry enhancement we also derive the
twisted sector   
\be Z_{-+} = {1\over 2}(\bar V_8 - \bar S_8) (O_8 - C_8)
\sum_{P=(p,\bar p) \in \Gamma} {q^{(p + v)^2/2} \bar q^{(\bar p + \bar
v)^2/2}\over
\tau_2^3|\eta (q)^8|^2 } \qquad ,
\label{twomp}
\ee and its projection
\be Z_{--} = -e^{\pi i V \cdot V}{1\over 2}(\bar V_8 - \bar S_8) ( O_8 + C_8)
\sum_{P=(p,\bar p) \in \Gamma} e^{2\pi i V \cdot P} {q^{(p + v)^2/2} \bar
q^{(\bar p + \bar v)^2/2}\over \tau_2^3|\eta (q)^8|^2 } \qquad .
\label{twomm}
\ee 
As for the CHL model discussed in Section 3, a sensible projection requires
$V
\cdot V$ to be  integer and $V \cdot P$ to be half-integer. In the twisted
sector, one finds massless states from the {\it tachyonic} $O_8$ factor when
$\bar p=0$ and $p^2=1$, while the co-spinor $C_8$ contributes massless states
only in the anti-decompactification limit $R_i\rightarrow 0$. Maximal
supersymmetry, together with its 32 supercharges, is restored in this limit, as
well as in the decompactification limit $R_i\rightarrow
\infty$, where the spinor $S_8$ contributes massless states in the untwisted
sector. Since in order to map the limit $R_i\rightarrow 0$ into the limit
$R_i\rightarrow
\infty$ one has to perform a T-duality that flips the chirality of
the spacetime spinors, \ie $C_8 \leftrightarrow S_8$, one retrieves the
original type IIB (4,4) model in both limits. In fact the self-dual
point is the point where the generalized HFK mechanism takes
place. Similar conclusions would have been reached had we started from the type
IIA (4,4) superstring.  This means that type IIA and type IIB (4,4) strings have
disconnected moduli spaces because at least perturbatively -- and the absence
of D-brane for any finite volume drastically reduces the possible sources of
non-perturbative effects -- one cannot obtain one 10D theory in any limit
point in the moduli space of the other! Moreover T-duality is part
of the gauge group of these theories much in the same way as in more familiar
heterotic compactifications \cite{nsw}.

The last remarks make the type II (4,0) strings a rather interesting arena to
test various duality conjectures. In particular, the gauge group is generically
abelian, in fact $U(1)^{(d+d)}$ in the model discussed above, and gets enhanced
only at points where $\bar p=0$ and
$p^2=1$. This typically correspond to free fermionic points. The enhanced
gauge symmetry is determined by the structure constants that appear in the
cubic supercurrent 
\cite{abk,dkv,lls}. Some type II (4,0) models admit U-dual type II (2,2)
models \cite{sv, kk}. In particular two classes of models in $D=4$ with
$N_V=4$ and $N_V=6$ seem to fall into this category \cite{sv,kk}. For the former
it seems a rather difficult task to construct a {\it perturbative}
type I dual while for the latter, as promised in the Introduction, one can
construct type I models that display exactly the same massless spectrum at
generic points of the moduli space. They correspond to toroidal
compactifications of the type I superstring with an unconventional Klein bottle
projection that does not allow for the introduction of D9-branes and their
open-string excitations
\cite{gepner,dp,park}. Indeed considering for simplicity tori with diagonal
metric one may take advantage of some arbitrariness in the $\Omega$-projection
and put
\be  {\cal K} = {1\over 2}\left({V_8 - S_8 \over \tau_2^{D/2} \eta(q\bar
q)^8} \right) \prod_{i=1, d}\sum_{m_i} e^{i\pi \delta_i m_i}(q\bar q)^{m_i^2
\over 2 R_i^2}
\label{unconventional}
\ee where $\delta_i=0,\pm 1$ are not fixed by the crosscap constraint
\cite{fpss}.  The resulting crosscap-to-crosscap amplitudes reads 
\be
\widetilde{\cal K} = {2^{D/2}\over 2}\left({V_8 - S_8 \over \tau_2^{D/2}
\eta(q\bar q)^8} \right) \prod_{i=1, d}\sum_{n_i} (q\bar q)^{(2n_i + \delta_i)^2
R_i^2/2} \qquad .
\ee  When at least one $\delta_i\neq 0$, no massless closed-string states flow
in the transverse channel. The O9-planes do not carry R$\times$R charge and
there is no room for introducing D9-branes and open-strings \cite{dp,gepner}.

As argued in the Introduction self-duality of the type IIB superstring in $D=10$
and application of the adiabatic argument
\cite{vwdual} to the type IIB superstring on tori with shifts $\sigma_V$ allow
one to relate type II (4,0) models to type I superstrings without open
strings. Indeed, it is easy to check that on the massless spectrum
$\Omega=S^{-1}(-)^{F_R} S$. Although the argument does not work in $D=10$ --
the $(-)^{F_R}$-projection gives the  type IIA superstring that is not S-dual to
the type I superstring obtained via the
$\Omega$-projection -- the argument seems to work in lower dimension, when
combined with a non-trivial $Z_2$-action on the compactification manifold, in
the case under consideration an order two shift $\sigma_V$ on a torus
$T^d$. Moreover the different inequivalent choices of $\sigma_V$ with $v=\bar
v$ are in one to one correspondence with the different inequivalent choices of
$\delta_i=v_i/2$ in (\ref{unconventional}). This allows one to deduce a precise
map between the two classes of models and conclude that they give rise to several
disconnected components of the moduli spaces of compactifications with
the same $N_V=d$ but with different global identifications. In particular, the
surviving T-dualities are those that commute with the shift $\sigma_V$.

Although the type II (4,0) strings and type I strings without open strings have
generically the same low-energy limit and share the same moduli space,
identifying the ones with the others requires a precise map between the
BPS\footnote{Here, BPS stands for Bogomolny-Prasad-Sommerfield.} states in the
two theories. In  particular, one should study the behavior of the type I
theory at images of the points of enhanced symmetry in the type II (4,0)
strings. The analysis of \cite{pw} suggests that the type I string coupling
diverge at these points and the relevant D-brane states become massless. It is
not clear however whether the R$\times$R charges of these states are quantized in
units of the fundamental R$\times$R charge of conventional D-branes.

\section{Final Comments}

Superstring vacua with 16 supercharges provide us with a plethora of models
for which tests of various duality conjectures seem both feasible and
interesting.  The advantage of this class of theories is that supersymmetry
alone does not fix uniquely the massless spectrum, but it is still powerful
enough to fix completely the low-energy lagrangian once the spectrum is fixed. In
particular, up to global
identifications, the moduli space parameterized by the scalars in the vector
multiplets is $O(d, d+n) / O(d)\times O(d+n)$, where $n+d=N_V$ is the
total number of vector multiplets. For {\it unconventional} type I
compactifications we have shown that
$n$ can take the values $n=16\times 2^{-b/2}$, with $b=0,2,4,6$, and $n=0$. 

The state of the art suggests that the existence of several disconnected
components can find a very simple rationale in terms of BPS unorientifolds,
\ie type I toroidal compactifications with a quantized 
NS$\times$NS antisymmetric tensor \cite{bpstor}. Other components, that
include type I theories with the minimum number of vector multiplets,
\ie $n=0$ or $N_V=d=10-D$, admit at least two different realizations. The first
in terms of type II superstrings without D-branes, \ie type II (4,0) models, the
second in terms  of type I
superstrings without open strings. An adiabatic argument suggests that the two
description are dual to one another. We have not succeeded in finding a way to
reduce the number of vectors in the unoriented closed-string spectrum of the
type I superstring to reproduce the disconnected components with
$N_V < 6$ in $D=4$ and
$N_V < 4$ in $D=6$, but we hope to investigate this issue in the future.

For type I superstrings other disconnected components of the moduli space of
vacua can be found by adding {\it open-string discrete torsion} and/or Wilson
lines that do not lie in $SO(32)$ \cite{bstw,bpstor}. In view of the
conjectured heterotic - type I duality this seems to lead to inconsistencies at
the non-perturbative level, because some non perturbative D-brane states are
expected to carry spinorial charges of $SO(32)$ and thus would not admit a
sensible action of $O(32)$ Wilson lines. However, it may well be possible
that in these component of the moduli space the type I superstrings do not
admit perturbative heterotic duals but rather type II (4,0) superstring duals.
Once again, establishing connections and checking various dualities for these
unconventional models seems to be more than compelling.

The recent insights in the duality realm of super Yang-Mills theories have
taught us that gauge symmetry is not a good {\it order parameter} for
discriminating between different theories. It simply represent a redundancy in
the description of  a system. Indeed, in different regimes the same {\it exact}
theory can display different gauge symmetries \cite{sw}\footnote{Clearly, in the
regime in which a gauge symmetry seems to have disappeared, the global symmetry
corresponding to gauge transformations with constant parameters is trivially
represented because the new composite degrees of freedom are expected to be
singlets of the original symmetry!}. Rolling among topologically different
superstring vacua with 8 supercharges, such as Calabi-Yau
compactifications of the type II superstrings or their heterotic and type I
duals, is the superstring counterpart of the above field-theory phenomenon. For
superstring vacua with 16 supercharges a mechanism analogous to 
black hole condensation in conifold transitions or to shrinking of 
instantons in transitions with tensionless strings has not been proposed so far.
Rolling among topologically different superstring vacua with 16 supercharges
seems forbidden or at least strongly suppressed. 
 
As a final remark, the present
development in our understanding  of superstrings
at the non-perturbative level makes the problem of vacuum selection sharper
and more compelling \cite{sei}. Superstring vacua with 16 supercharges include
vacua that do not admit a natural geometric interpretation such as
unconventional un-orientifolds and asymmetric orbifolds that are clearly
calling for a thorough analysis. The existence of an underlying theory of
membranes and/or pentabranes bringing into play new consistent {\it
non-perturbative} vacuum configurations, corresponding to the wrapping branes
of complicated topology around cycles of the compactification manifold, can
only make the problem more severe at first glance.

\section{Note Added} While this work was being typed, I was informed by 
A. Sagnotti that E. Witten \cite{wittor} was also considering issues related to
the quantized NS$\times$NS antisymmetric tensor introduced in \cite{bpstor}.

\section{Acknowledgements} I would like to acknowledge useful discussions
with C. Angelantonj, D. Polyakov, G. Pradisi, K. Ray, Ya. Stanev, and
A. Sagnotti, whom I especially thank for carefully reading the manuscript. I
would like to thank the  organizers of the ``Abdus Salam Memorial Meeting" for
their kind hospitality at the Abdus Salam ICTP while this work has been brought
to completion.


\begin{thebibliography}{99}

\bibitem{bpstor} {M. Bianchi, G. Pradisi and A. Sagnotti, {\it Nucl. Phys.} {\bf
B376} (1991) 365.}
\bibitem{dix}{L. Dixon, in {\it Proceedings  of the ICTP Summer Workshop on
High Energy Physics and Cosmology}, Trieste, Italy, 1987 (World Scientific,
Singapore).} 
\bibitem{send}{See \eg A. Sen, hep-th/9802501, for a comprehensive review.}
\bibitem{as} {A. Sagnotti, in {\it ``Non-Perturbative Quantum  Field
Theory''}, eds. G. Mack et al (Pergamon Press, 1988), p. 521.}
\bibitem{bssys} {M. Bianchi and A. Sagnotti, {\it Phys. Lett.} {\bf B247}
(1990) 517.}
\bibitem{cjp} {J.~Polchinski, hep-th/9510017; E. Gimon and J. Polchinski,
hep-th/9601038; S. Chaudhuri, C. Johnson and J.~Polchinski,  hep-th/9602052.} 
\bibitem{dp} {A. Dabholkar
and J. Park, hep-th/9604178; E. Gimon and C.V. Johnson, hep-th/9606176.}
\bibitem{gepner} {C. Angelantonj, M. Bianchi, G. Pradisi, A. Sagnotti and Ya.S.
Stanev, {\it Phys. Lett.} {\bf B387} (1996) 743.}
\bibitem{bstw} {M. Bianchi and A. Sagnotti, {\it Nucl. Phys.} {\bf B361} (1991)
519.}
\bibitem{vwdt} {C. Vafa and E. Witten, hep-th/9409188.}
\bibitem{sese} {A. Sen and D. Sethi, hep-th/9703157.}
\bibitem{bpsmin} {M. Bianchi, G. Pradisi and A. Sagnotti, {\it Phys. Lett.}
{\bf B273} (1991) 389.}
\bibitem{fpss} {D. Fioravanti, G. Pradisi and A. Sagnotti, {\it Phys. Lett.}
{\bf B321} (1994) 349; G. Pradisi, A. Sagnotti and Ya.S. Stanev, {\it Phys.
Lett.} {\bf B354} (1995) 279; {\bf B356} (1995) 230. {\bf B381} (1996) 97.}
\bibitem{chl} {S. Chaudhuri, G. Hockney and J. Lykken, hep-th/9505054.}
\bibitem{sv} {A. Sen and C. Vafa, hep-th/9508064; A. Kumar, hep-th/9601067.}
\bibitem{kk} {E. Kiritsis and C. Kounnas, hep-th/9703059; A. Gregori, E.
Kiritsis, C. Kounnas, N. Obers, P. Petropoulos and B. Pioline, hep-th/9708062.}
\bibitem{nsv} {K. Narain, M. Sarmadi and C. Vafa, {\it Nucl. Phys.} {\bf B288}
(1987) 551; L. Dixon, J. Harvey, C. Vafa and E. Witten, {\it Nucl. Phys.} {\bf
B261} (1985) 678; {\bf B274} (1986) 285.}
\bibitem{cp} {S. Chaudhuri and J. Polchinski, hep-th/9506048.}
\bibitem{witdyn} {E. Witten, {\it Nucl. Phys.} {\bf B443} (1995) 85; C.~Hull and
P.~Townsend, {\it Nucl. Phys.} {\bf B438} (1995) 109; {\it ibid.} 
{\bf B451} (1995) 525.} 
\bibitem{dht} {A. Dabholkar, {\it Phys. Lett.} {\bf B357} (1995) 307;  C. Hull,
{\it Phys. Lett.} {\bf B357} (1995) 545; A. Tseytlin, {\it Phys. Lett.} {\bf
B367} (1996) 84;  {\it Nucl. Phys.} {\bf B467} (1996) 383.}
\bibitem{pw} {J. Polchinski and E. Witten, {\it Nucl. Phys.} {\bf B460} (1996)
525.}
\bibitem{lmst} {W. Lerche, R. Minasian, C. Schweigert and S. Theisen,
hep-th/9711104.}
\bibitem{ap}{A. Polyakov, hep-th/9711002.} 
\bibitem{salsez}{A. Salam and E. Sezgin, {\it Supergravities in Diverse
Dimensions}, Chapter 2, North Holland / World Scientific 1989.}
\bibitem{nsw}{K. Narain, {\it Nucl. Phys.} {\bf B169} (1987) 41; K. Narain, M.
Sarmadi and E. Witten, {\it Nucl. Phys.} {\bf B279} (1987) 369.}
\bibitem{dff} {L. Andrianopoli, R. D'Auria, S. Ferrara, P. Fr\'e, M. Trigiante,
hep-th/9611014; {\it idem} and R. Minasian hep-th/9612202.}
\bibitem{chiral} {C. Angelantonj, M. Bianchi, G. Pradisi, A.
Sagnotti and Ya.S. Stanev, {\it Phys. Lett.} {\bf B385} (1996) 96.}
\bibitem{vwdual} {C. Vafa and E. Witten, hep-th/9507050.}
\bibitem{senk} {A. Sen, hep-th/9504027; J. Harvey and A.
Strominger, hep-th/9504047;  M.~Duff, hep-th/9501030.} 
\bibitem{asp} {P. Aspinwall, hep-th/9508154; hep-th/9611137.}
\bibitem{ss} {J. Schwarz and A. Sen, hep-th/9507027.}
\bibitem{cl} {S. Chaudhuri and D. Lowe, hep-th/9508144; hep-th/9512226.}
\bibitem{abk} {I. Antoniadis, C. Bachas, K. Kounnas; {\it Nucl. Phys.} {\bf
B289} (1987) 87; H. Kawai, J. Lewellen, H. Tye, {\it Nucl. Phys. } {\bf B288}
(1987) 1; I. Antoniadis and C. Bachas; {\it Nucl. Phys.} {\bf B298} (1988)
586.}
\bibitem{fk} {S. Ferrara and K. Kounnas, {\it Nucl. Phys.} {\bf B328} (1989)
406.}
\bibitem{dkv} {L. Dixon, V. Kaplunovsky and C. Vafa, {\it Nucl. Phys.} {\bf
B294} (1987) 43.}
\bibitem{lls} {W. Lerche, D. L\"ust and A. Schellekens, {\it Phys.Rep.} {\bf 177}
(1989)  1-140.}
\bibitem{mt} {J.~Schwarz, hep-th/9510086; M. Duff, {\it Int. J. Mod. Phys.} 
{\bf 11} (1996) 5623.} 
\bibitem{horwit} {P. Ho\v{r}ava and E.~Witten, hep-th/9510209.}
\bibitem{gsw} {M.~Green, J.~Schwarz and E.~Witten, {\it Superstring Theory},
Cambridge University Press, 1987.}
\bibitem{park} {J. Park, hep-th/9611169.}
\bibitem{kr} {K. Dasgupta and S. Mukhi, hep-th/9512219; A. Sen, hep-th/9602010;
A. Kumar and K. Ray,  hep-th/9602144.}
\bibitem{ft} {C. Vafa, hep-th/9602022; D.R. Morrison and C. Vafa,
hep-th/9602114, hep-th/9603161.}
\bibitem{senf} {A. Sen, hep-th/9605150; hep-th/9709159.}
\bibitem{dm} {K. Dasgupta and S. Mukhi, hep-th/9606044; S. Mukhi,
hep-th/9710004.}
\bibitem{blum} {J. Blum and A. Zaffaroni, hep-th/9607019; J. Blum,
hep-th/9608053 ; Gopakumar and S. Mukhi, hep-th/9607057.}
\bibitem{sengp} {A. Sen, hep-th/961186; hep-th/9702061.}
\bibitem{tutti} {M. Berkooz, R.G. Leigh, J.Polchinski, J.H. Schwarz, N.
Seiberg and E. Witten,  hep-th/9605184.}
\bibitem{kak} {Z. Kakushadze, hep-th/9704059.}
\bibitem{twelve} {M. Bianchi, S. Ferrara, G. Pradisi, A. Sagnotti and Ya.S.
Stanev, {\it Phys. Lett.} {\bf B387} (1996) 64.}
\bibitem{bpzfms}{A. Belavin, A. Polyakov and A. Zamolodchikov, {\it Nucl. Phys.}
{\bf B241} (1984) 333; D. Friedan, E. Martinec and S. Shenker, {\it Nucl. Phys.}
{\bf B271} (1986) 93.}
\bibitem{aa}{A. Alessandrini, {\it Nuovo Cim.} {\bf 2A} (1971) 321; 
A. Alessandrini and D. Amati, {\it Nuovo Cim.} {\bf 4A} (1971) 793.}.
\bibitem{gs}{M. Green and J. Schwarz, {\it Phys. Lett.} {\bf 149B} (1984) 117;
{\it ibid.} {\bf 151B} (1985) 21.}
\bibitem{ghmr}{D. Gross, J. Harvey, E. Martinec and R. Rohm, {\it Nucl. Phys.}
{\bf B256} (1985) 253.}
\bibitem{ms} {N. Marcus and A. Sagnotti, {\it Phys. Lett.} {\bf B188} (1987)
58; M. Bianchi and A. Sagnotti, {\it Phys. Lett.} {\bf B211} (1988)
407.}
\bibitem{bsrmg} {M. Bianchi and A. Sagnotti, {\it Phys. Lett.} {\bf B231}
(1990) 389.}
\bibitem{ps}{G. Pradisi and A. Sagnotti, {\it Phys. Lett.} {\bf B216} (1989)
59.} 
\bibitem{jc}{J. Cardy, {\it Nucl. Phys.} {\bf B324} (1989) 581.}
\bibitem{ev}{E. Verlinde, {\it Nucl. Phys.} {\bf B300} (1988) 360.}
\bibitem{pc} {Y. Cai and J. Polchinski, {\it Nucl. Phys.} {\bf B376} (1991)
365.}
\bibitem{swtach} {C. Vafa, {\it Nucl. Phys.} {\bf B273} (1986) 592; L. Dixon, J.
Harvey, {\it Nucl. Phys.} {\bf B274} (1986) 93; N. Seiberg and E. Witten, 
{\it Nucl. Phys.} {\bf B276} (1986) 272;  L. Alvarez-Gaum\'e, G. Moore and C.
Vafa, {\it Comm. Math. Phys.} {\bf 106} (1986) 1.}
\bibitem{bd} {O. Bergman and M. Gaberdiel, hep-th/9701137;
J. Blum and K. Dienes, hep-th/9707160; hep-th/9707148.}  
\bibitem{mbik} {M. Bianchi, {\it Open Strings and Dualities},
hep-th/9712020, ROM2F-97/52, talk delivered at the {\it V Italo-Korean Meeting 
on Relativistic  Astrophysics}, Seoul-Suanbo, 1-7 September, 1997.} 
\bibitem{hor} {P. Ho\v{r}ava, {\it Nucl. Phys.} {\bf B327} (1989) 461, {\it
Phys. Lett.}  {\bf B231} (1989) 251.}
\bibitem{gep} {D. Gepner, {\it Lectures on ${\cal N}=2$ String Theory}, Spring
School on Superstrings, A. Salam ICTP, Trieste, 3-11 April, 1989.} 
\bibitem{sw} {N. Seiberg and E. Witten,  {\it Nucl. Phys.} {\bf B426} (1994)
19;   {\it ibid.} {\bf B431} (1994) 484; N. Seiberg, hep-th/9411149.}
\bibitem{sei} {N. Seiberg, {\it The Superworld}, talk delivered at the {\it
Abdus Salam Memorial Meeting}, A. Salam ICTP, Trieste, 19-21 November, 1997.}
\bibitem{wittor}{E. Witten, hep-th/9712028.}


\end{thebibliography}
\end{document}